\pdfoutput=1 
\documentclass[conference, a4paper]{IEEEtran}
\IEEEoverridecommandlockouts
\usepackage{graphicx}
\usepackage{cite}
\usepackage{amsmath,amssymb,amsfonts}
\usepackage{algorithmic}
\usepackage{textcomp}
\usepackage{xcolor}
\usepackage{hyperref}
\usepackage[
  separate-uncertainty = true,
  multi-part-units = repeat
]{siunitx}

\def\BibTeX{{\rm B\kern-.05em{\sc i\kern-.025em b}\kern-.08em
    T\kern-.1667em\lower.7ex\hbox{E}\kern-.125emX}}
\begin{document}
\begin{titlepage}
	\begin{center}
		
		\Huge
		\textbf{The Use of MEMS Accelerometers for Remote Activity and Living Parameters Monitoring							
		}
		
		\vspace{0.5cm}
		\LARGE
		Accepted version
		
		\vspace{1.5cm}
		
		\text{Marcin Kolakowski, Natalia Osiadala}
		
		\vspace{.5cm}
		\Large
		Institute of Radioelectronics and Multimedia Technology
		
		Warsaw University of Technology
		
		Warsaw, Poland,
		
		contact: marcin.kolakowski@pw.edu.pl

		\vspace{2cm}

	\end{center}
	
	\Large
	\noindent
	\textbf{Originally presented at:}
	
	\noindent
2020 Baltic URSI Symposium (URSI), Warsaw, Poland, 2020
	
	\vspace{.5cm}
	\noindent
	\textbf{Please cite this manuscript as:}
	
	\noindent
N. Osiadala and M. Kolakowski, "The Use of MEMS Accelerometers for Remote Activity and Living Parameters Monitoring," 2020 Baltic URSI Symposium (URSI), Warsaw, Poland, 2020, pp. 73-76, doi: 10.23919/URSI48707.2020.9254025.
	
	\vspace{.5cm}
	\noindent
	\textbf{Full version available at:}
	
	\noindent
	\url{https://doi.org/10.23919/URSI48707.2020.9254025}

	%		\vspace{.5cm}
	%		\noindent
	%		\textbf{Additional information:}
	%		
	%		\noindent
	%		The dataset used in the study is available at Zenodo:
	%		
	%		\noindent
	%		Marcin Kolakowski. (2021). UWB Channel Impulse Responses Registered in a Furnished Apartment (Version 1.0) [Data set]. Zenodo. \url{http://doi.org/10.5281/zenodo.4742391}
	
	\vfill
	
	\large
	\noindent
	© 2020 IEEE. Personal use of this material is permitted. Permission from IEEE must be obtained for all other uses, in any current or future media, including reprinting/republishing this material for advertising or promotional purposes, creating new collective works, for resale or redistribution to servers or lists, or reuse of any copyrighted component of this work in other works.
\end{titlepage}

\title{The Use of MEMS Accelerometers for Remote Activity and Living Parameters Monitoring
\thanks{This work was supported by the National Centre for Research and Development, Poland under Grant AAL/Call2016/3/2017 (IONIS project) and the Foundation for the Development of Radiocommunication and Multimedia.}
}

\author{\IEEEauthorblockN{Natalia Osiadala}
\IEEEauthorblockA{
\textit{Warsaw University of Technology}\\
\textit{Institute of Radioelectronics and Mult. Tech.} \\
Warsaw, Poland \\
n.osiadala@stud.elka.pw.edu.pl}
\and
\IEEEauthorblockN{Marcin Kolakowski}
\IEEEauthorblockA{
\textit{Warsaw University of Technology}\\
\textit{Institute of Radioelectronics and Mult. Tech.} \\
Warsaw, Poland \\
m.kolakowski@ire.pw.edu.pl}
}

\maketitle
\begin{abstract}
In the paper a ballistocardiographic sensor for remote monitoring of activity and vital parameters is presented. The sensor is mainly intended for use in monitoring systems supporting care of older people. It allows to detect occupancy of a piece of furniture, to which it is attached and to estimate basic vital parameters (heart and respiration rates) of the monitored person. The presented device includes  three inertial sensors: two accelerometers of different parameters and prices and one reference BCG module. The device sends the measurement results to the external server over WiFi. The vital parameters are estimated based on the Continuous Wavelet Transform of the registered acceleration signals. User's presence is detected by tracking changes in acceleration measured in axes parallel to the ground.
\end{abstract}


\section{Introduction}

European population is aging fast. According to the report by the European Commission \cite{europeancommission2018AgeingReport2017a}, the share of people aged over 65 years is expected to increase from 18 percent in 2013 to 28 percent in 2060. As over the years, the human body is becoming more and more susceptible to various diseases, providing such a large number of elderly people with a safe and comfortable life will be one of the most important challenges facing our societies.
Older people often suffer from diseases (e.g. dementia), which make them require constant care and monitoring. The research has shown that, from their point of view, it would be most beneficial to provide such care in their own homes rather than long-term care facilities. The progressing development of technology, like the Internet of Things, opens new possibilities of using advanced systems allowing for remote monitoring of activity and health assessment of  elderly persons.

Most of the currently available vital parameters monitoring solutions require sensors are devices worn by a monitored person (smartwatches, armbands). Those solutions, due to inconveniences related to wearing them,  often meet with a lack of user’s acceptance. Therefore there is a need for development of non-intrusive ways of monitoring and diagnosis.

\begin{figure}[b]
\centering
\centerline{\includegraphics[width=0.6\linewidth]{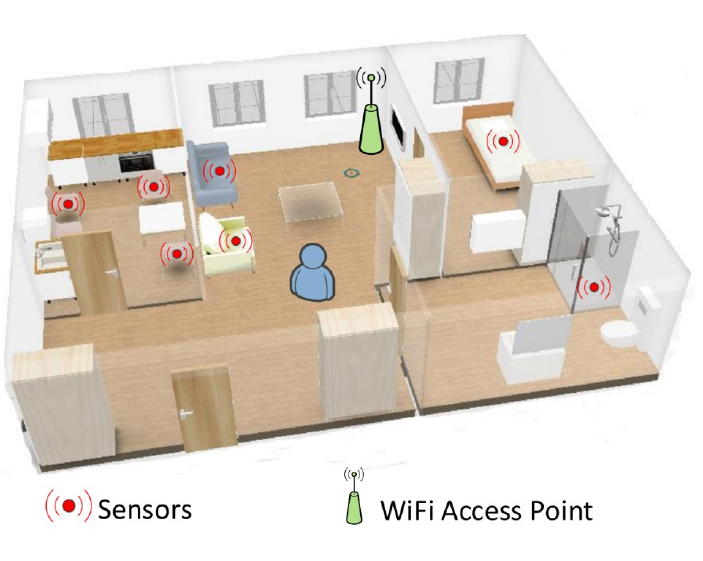}}
\caption{A concept of an older people activity monitoring system}
\label{fig:monitoring_system}
\end{figure}

The concept of an exemplary system for non-intrusive, remote monitoring of elderly person’s activity at his/her home is shown in Fig. \ref{fig:monitoring_system}.

The presented system consists of a network of sensors attached to various pieces of furniture, with which a person has frequent and long-term contact (e.g. armchair, chair, bed) and a Wi-Fi access point. The sensors allow to locate the person in specific areas by detecting occupancy of particular pieces of furniture. In this type of solutions, versatility of the sensors and a broad scope of their functions are important. It would be desired if the sensors, apart from occupancy detection, allowed to estimate vital parameters of the monitored people. Those two needs can be fulfilled by employing ballistocardiographic senors.
The ballistocardiography is a method consisting in  vibrations caused by body movements due to  heart contractions and respiration. Ballistocardographic sensors are usually attached to furniture pieces, which are occupied by the monitored person for a significant amount of time (e.g. beds, armchairs), and register the vibrations propagating through their frames. Based on the registered acceleration signals, it is possible to determine pulse and respiratotion rates.

In the literature, many examples of ballistocardiographic sensors can be found. Most of them include load-cell sensors placed under the bed legs \cite{leePhysiologicalSignalMonitoring2016} or under the mattress \cite{albukhariBedEmbeddedHeartRespiration2019}. Although these solutions allow for precise determination of vital parameters, they are either difficult to install or too expensive to use in systems which would be affordable for the majority of the elderly. More cost effective solutions could be achieved by using sensitive accelerometers as in \cite{heEarwornContinuousBallistocardiogram2012} and \cite{jhne-radenSignalDetectionAccuracy2017}.

There are also several ballistocardiographic sensors available on the market among which are EmFit \cite{EMFITSleepTracking}, which is a ferro-electret sensor mat placed under the mattres and Murata SCA11H \cite{SCA11HContactlessBed} including a low noise accelerometer.

In order to estimate vital parameters of the monitored person, the registered vibration signals are processed using various methods. One of the popular methods is signal filtration. The method utilizes the fact that the frequency components of the registered signals corresponding to vital functions are located in specific bands: respiration in 0--0.5 Hz and heartbeat in 1--25 Hz \cite{albukhariBedEmbeddedHeartRespiration2019}. A method employing low-pass and band-pass filtering is presented in \cite{zinkUnobtrusiveNocturnalHeartbeat2017}. Another example of an efficient method is Continuous Wavelet Transform (CWT)  \cite{gilaberteHeartRespiratoryRate2010}.

The following paper presents a ballistocardiographic sensor device, in which three inertial sensors are used:  two models of accelerometers and a reference BCG module. The device enables registration of acceleration and remote transmission of measurement results via Wi-Fi network. The performed research was aimed at determining the suitability of the selected components for ballistocardiographic measurements, especially to see, whether a less expensive accelerometer can be successfully used.
\section{Device description}
\label{sec:device}
\subsection{Device design}

\begin{figure}[t]
\centering
\centerline{\includegraphics[width=\linewidth]{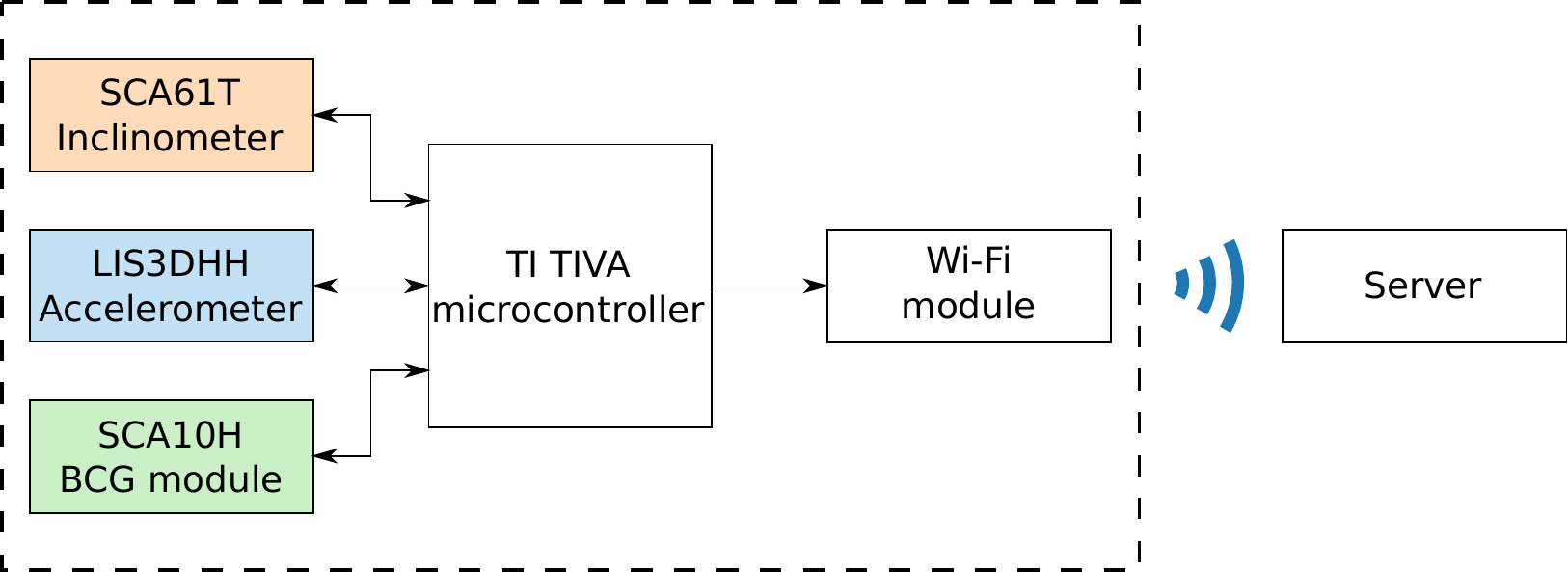}}
\caption{The block diagram of the sensor}
\label{fig:device}
\end{figure}

The block diagram of the proposed ballistocardiographic sensor is shown in Fig. \ref{fig:device}. 

The proposed sensor consists of two accelerometers, a BCG  module, a microcontroller and a Wi-Fi module. The intended use of the device assumes fixing it to a furniture frame or placing close to the monitored person and registering vibrations caused by the movements of the body related to breathing and heartbeat, which propagate through the furniture.

The LIS3DHH \cite{stmicroelectronicsMEMSMotionSensor2017} accelerometer and SCA61T \cite{murataSCA61TInclinometerSeries2015} inclinometer record those small movements at a 100 Hz rate and pass the results to the Texas Instruments Tiva microcontroller over SPI. The reference BCG module SCA10H processes measured acceleration on its own and sends the obtained vital parameters to the microcontroller using UART.  The collected results are sent over Wi-Fi to the computer and are processed there. If the presence of the monitored person is detected, the vital parameters are estimated.

The inertial sensors used in the system come from two different price groups. The market price of the LIS3DHH accelerometer fluctuates around ten dollars, whereas the SCA61T  inclinometer is about four times more expensive. The latter is used in the Murata SCA10H BCG module, which was placed in the device to act as reference. Basic parameters of the employed sensors are presented in Table \ref{tab:accelerometers}.

The LIS3DHH sensor allows to measure acceleration in three axes, while SCA61T is nominally a single-axis inclinometer (but allows access to values measured in the second perpendicular channel). LIS3DHH has a wider range of measured acceleration changes and is more sensitive than SCA61T. However, it introduces higher noise to the measured values, which might be problematic when working with low amplitude signals.

\begin{table}[t]
\caption{Basic parameters of accelerometers}
\begin{center}
\begin{tabular}{|l|c|c|}
\hline
\textbf{Parameter}&\textbf{LIS3DHH \cite{stmicroelectronicsMEMSMotionSensor2017}}&\textbf{SCA61T \cite{murataSCA61TInclinometerSeries2015}}\\
\hline 
Range [g] & \SI{\pm 2.5}{} & \SI{\pm 1}{} \\
Bits&16&11\\ 
Sensitivity [mg/LSB]&0.076&1.22\\
Noise [$\mu$g/Hz]&45&14\\
Max. sampling frequency [kHz]&1.1&5.5\\
\hline
\end{tabular}
\label{tab:accelerometers}
\end{center}
\end{table}

\subsection{Data transmission}
\begin{figure}[b]
\centering
\centerline{\includegraphics[width=0.8\linewidth]{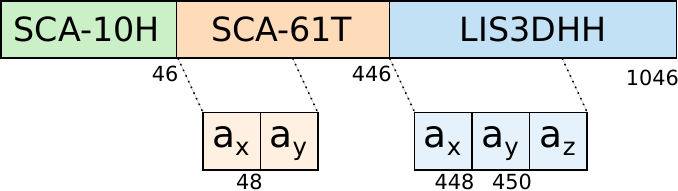}}
\caption{The format of the packet  sent to the server}
\label{fig:packet}
\end{figure}

The accelerometers perform measurements with frequency of 100 Hz, whereas the SCA10H module returns vital parameters once per second. The microcontroller puts the received results in packets, which structure is presented in Fig. \ref{fig:packet}.

The packet starts with  a 46-byte frame containing SCA10H results (incl. heart and respiration rates, occupancy). The remaining part of the packet  contains 100 samples from each accelerometer measured in two (SCA61T) and three (LIS3DHH) axes. Both accelerometers return measured values as two-byte variables. The total size of the transmitted packet is constant and equals 1046 bytes. The minimum transmission rate needed is therefore about \SI{8.37}{kbps}.

\section{Data processing}
\label{sec:processing}
\subsection{Occupancy detection}
\begin{figure}[t]
\centerline{\includegraphics[width=\linewidth]{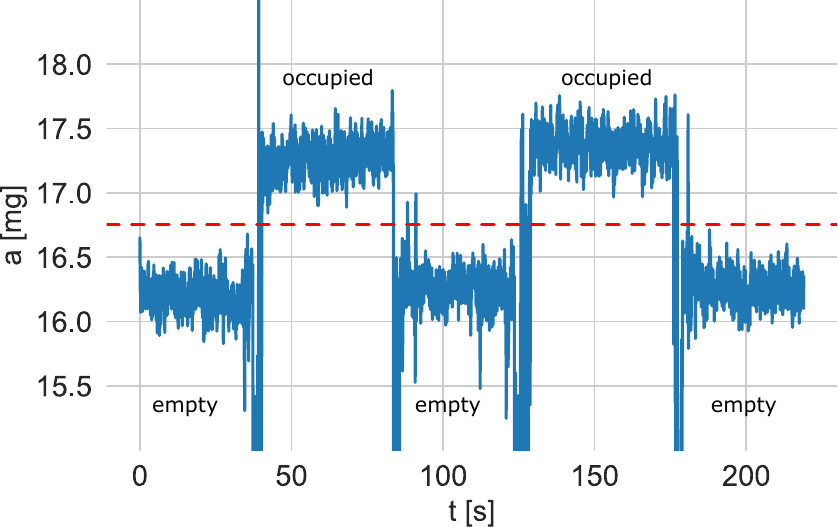}}
\caption{Acceleration measured in axis x (parallel to the ground) during bedding-down--getting-up sequence}
\label{fig:acc_occ}
\end{figure}

When the person takes a chair or beds down, he puts pressure on the furniture frame. Since this pressure is not evenly spread, the furniture and the attached sensor tilt, which causes changes in acceleration measured in the axes parallel to the ground. The acceleration signal registered during bedding-down--getting-up sequence is presented in Fig. \ref{fig:acc_occ}.

Bedding down causes an increase in the measured acceleration. When the persons gets up, the device tilts back to its initial state. In the proposed method, furniture occupancy is detected by comparing the measured acceleration to a threshold. The threshold value depends on the many factors e.g. users weight, furniture type, sensor placement and should be calibrated for each setup.

\subsection{Vital parameters analysis}
\begin{figure}[t]
\centerline{\includegraphics[width=\linewidth]{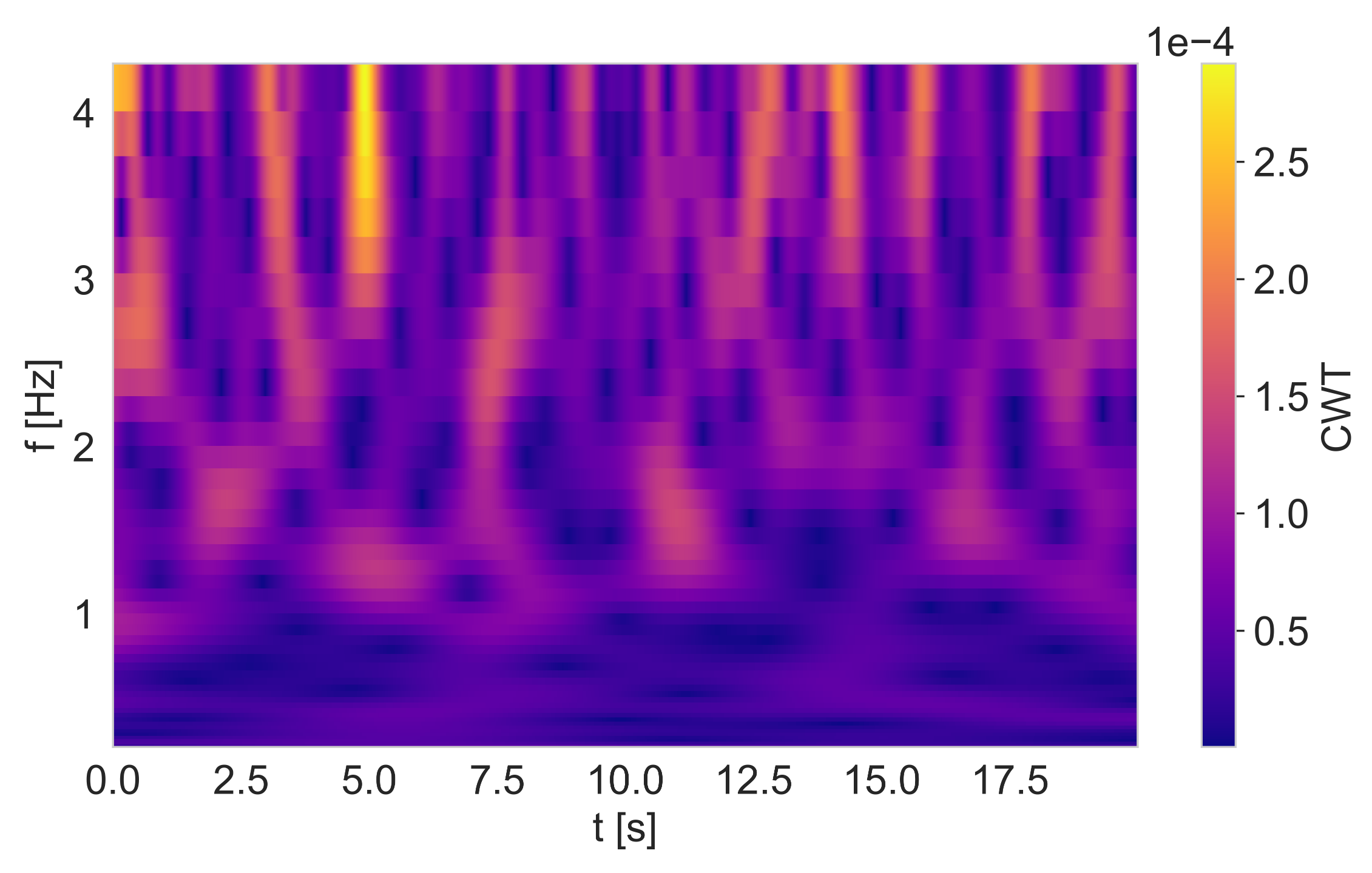}}
\caption{CWT scalogram for acceleration registered with LIS3DHH}
\label{fig:cwt_scalogram}
\end{figure}

In the conducted study, the vital parameters (heart rate, respiration rate) are estimated using Continuous-Wavelet-Transform-based (CWT) method. The CWT was calculated using Morlet wavelets, which shape are similar to acceleration changes caused by heart muscle contractions. An exemplary scalogram obtained for the results gathered with the LIS3DHH accelerometer is presented in Fig.\ref{fig:cwt_scalogram}.

\begin{figure}[t]
\centerline{\includegraphics[width=0.9\linewidth]{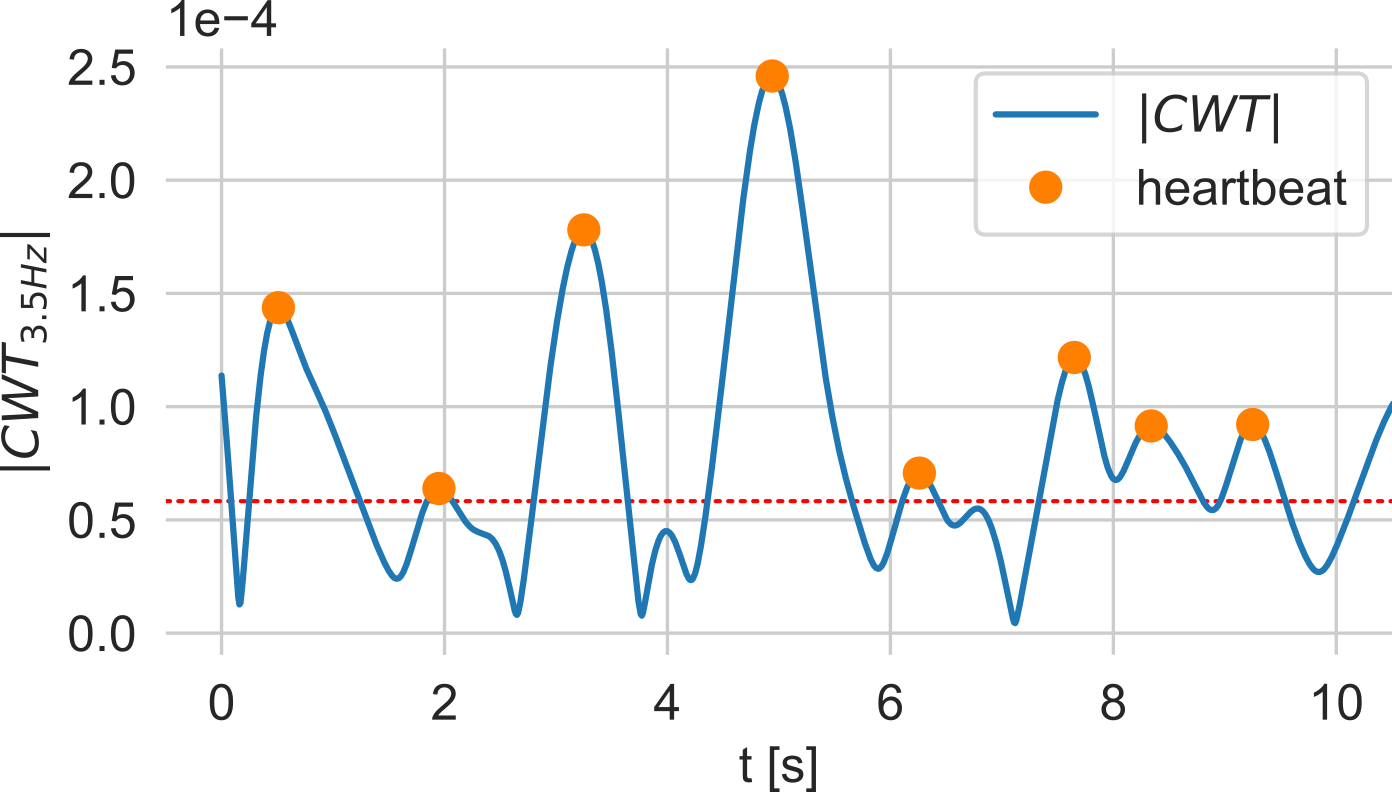}}
\caption{Module of the CWT for frequency \SI{3.5}{Hz} with heartbeat moments marked. The red line is a calibrated threshold.}
\label{fig:cwt_heart}
\end{figure}

The CWT shows signs of periodicity in the 1--4 Hz band, which is in the  1--25 Hz range associated with heartbeat inducted movements \cite{albukhariBedEmbeddedHeartRespiration2019}. The heartbeat and respiration rates can be determined through analysis of the CWT coefficients modules for different frequencies (3.5 and 0.8 Hz respectively). The changes of the CWT module for 3.5 Hz wavelet are presented in Fig. \ref{fig:cwt_heart}.

The peaks present in the CWT coefficient module signal, correspond to consecutive heart beats or breaths. The heartbeat and breathing rates can be estimated by calculating peaks' repetition period. In order to avoid errors caused by noisy measurements, the peaks above certain threshold are considered. In the proposed method the threshold value is calibrated by taking a measurement and calculating a 5th percentile of the peaks amplitude.   The heart and respiration rates are then calculated using the following dependency:
\begin{equation}
{\rm heart/respiration\,rate} = \frac{60}{\frac{1}{n}\sum_{i=0}^{n}{t_{i}}}\label{eq}
\end{equation}
where $t_{i}$ is a time between two consecutive peaks and $n$ is the number of peaks detected in the last minute.

\begin{figure*}
\centerline{\includegraphics[width=0.9\linewidth]{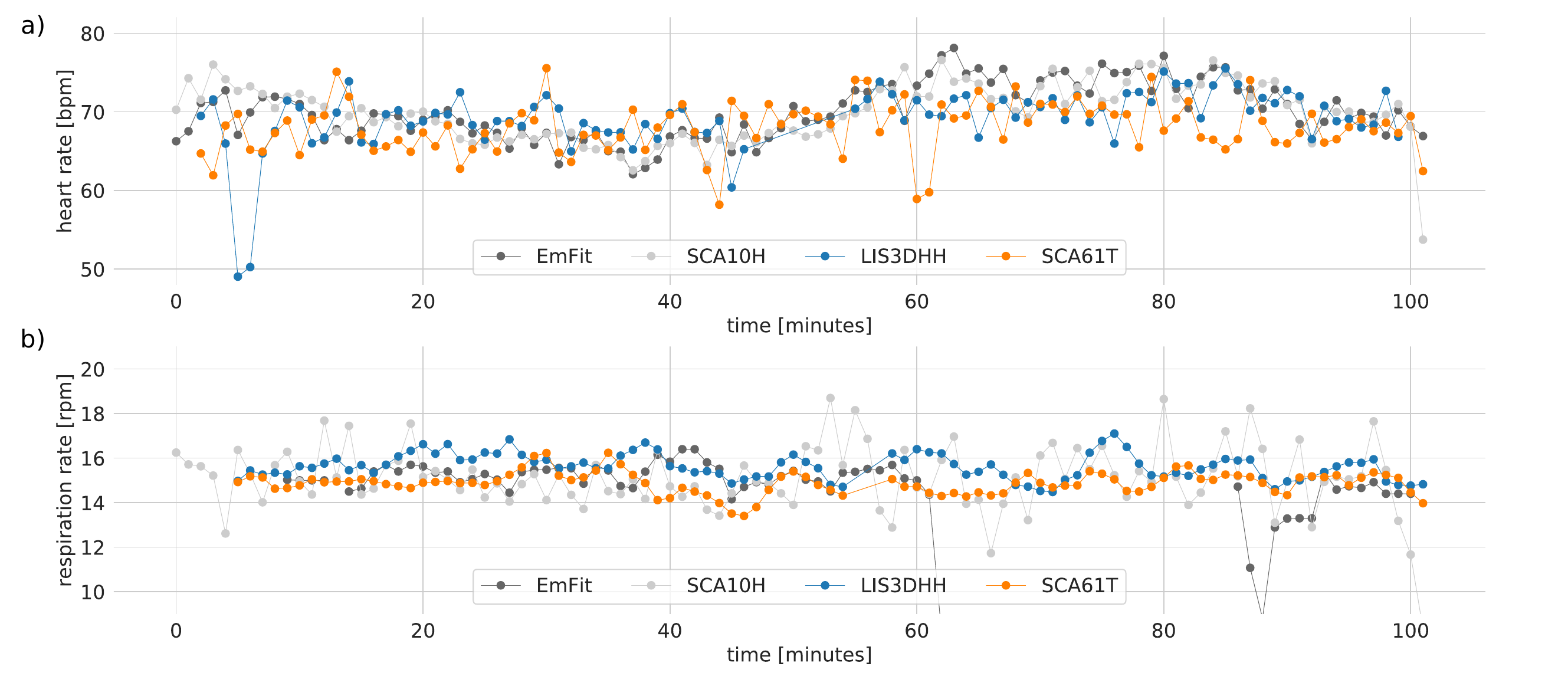}}
\caption{Estimated vital parameters (averaged per second): a) heart rate, b) respiration rate}
\label{fig:results}
\end{figure*}

\section{Experiments}
\label{sec:experiments}

The completed device was experimentally tested. The tests included detection of bed occupancy and vital parameters (heart rate, respiration rate) estimation. During the tests, the device was fixed to the bed frame near the monitored person. The obtained results were compared to ones obtained with reference BCG sensors: SCA10H, which is included in the device and EmFit QS (medically certified version) \cite{EMFITSleepTracking}.

\subsection{Occupancy detection}

\begin{figure}[t]
\centerline{\includegraphics[width=.75\linewidth]{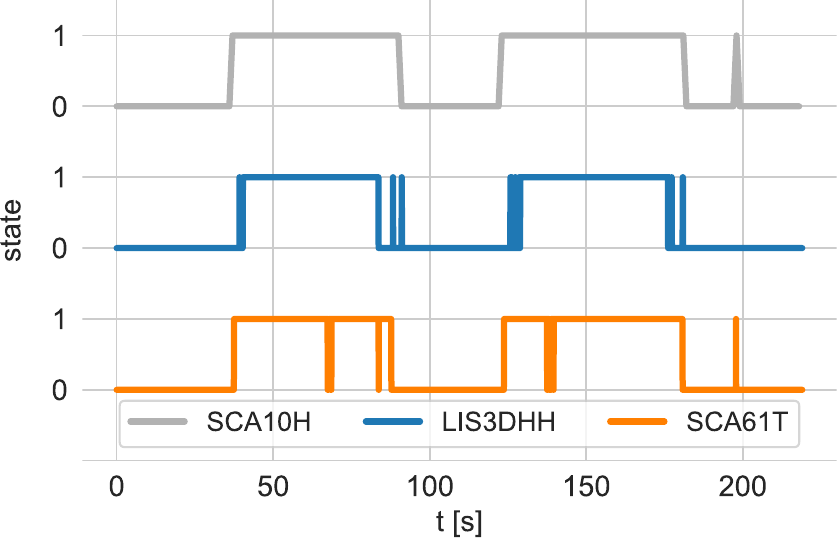}}
\caption{The results of bed occupancy detection (0 - empty, 1 - occupied)}
\label{fig:occ_res}
\end{figure}

The experiment consisted in bedding down and getting up two times (as presented in Fig. \ref{fig:acc_occ}. The results of occupancy detection using the proposed method are shown in Fig. \ref{fig:occ_res}.

The results achieved with the proposed sensor and the reference SCA10H were very similar. In case of LIS3DHH and SCA61T there were some situations, when the sensor detected empty bed, when it was occupied, but duration of those changes was short and they can be easily removed by averaging.

\subsection{Vital parameters analysis}

The second part of the study consisted in measurement of vital signs during a one-hour-forty-minute session when the monitored person was sleeping. The results of the heart and respiratory rates estimation are presented in Fig.\ref{fig:results}.

The obtained heart rates  are close to the ones measured with the reference sensors. For most of the time, the difference does not exceed 5 beats per minute. Given the difference in heart rates reported by the reference sensors SCA10H and EmFit, it is hard to tell, which value is accurate. All of the tested sensors allow to capture long-term trend of heart rate changes.

Due to small acceleration changes caused by breathing, measuring respiratory rate is not always possible. There are moments, where the constructed sensor and reference devices do not report any breathing. The rates obtained using both of the accelerometers are close to the ones measured with the EmFit sensor. The results reported by SCA10H have high variability, which puts its measurement reliability in doubt.

\section{Conclusions}
The paper presents an implementation of  a  ballistocardiographic sensor device for  monitoring the activity and vital parameters of an elderly person in his/her home. The conducted research has shown that the sensor allows to achieve the above goals. The results obtained using the proposed methods were similar to the ones returned by commercially available and recognized devices.

Based on experimental results, it can be concluded that it is possible to successfully use cost-effective accelerometer (e.g. LIS3DHH) in BCG sensor construction. It is an important remark, which would allow for construction of more affordable BCG sensors in the future. 

The proposed sensor samples the acceleration with a relatively low rate (100 HZ instead of typical 1kHz). The required  transmission rate is about \SI{8.37}{kbps}, which would allow to use less demanding radio interfaces such as Bluetooth.

%\section*{References}

\bibliographystyle{IEEEtran}
\bibliography{bcg_bib}

\end{document}